%% file: psb_bal_Submission_version.tex
\documentclass[usenatbib]{mn2e}
\usepackage[utf8x]{inputenc}
\usepackage{graphicx}
\usepackage{amssymb}
\usepackage{amsmath}
\usepackage{amssymb}
\usepackage{amsmath}
\usepackage{color}
\usepackage{times}
\usepackage{lineno}

\newcommand{\balnamelong}{DES QSO J033049.33-283249.7\textrm{}} 
\newcommand{\balnameshort}{DES QSO J0330-28\textrm{}}
\newcommand{\desfilters}{\emph{grizY}}
\newcommand{\vhsfilters}{\emph{JHK}}

\newcommand{\kms}{km s$^{-1}$\textrm{}}

\newcommand{\redchi}{$\chi^{2}_{\textrm{red}}$}

\title{Discovery of a z=0.65 Post-Starburst BAL Quasar in the DES Supernova Fields}
\author[Mudd et al.]{\parbox{18cm}{Dale Mudd$^{1}$, Paul Martini$^{1,2}$, Suk Sien Tie$^{1,2}$, Chris Lidman$^{3}$, Richard McMahon$^{4,5}$, Manda Banerji$^{4,5}$, Tamara Davis$^{6}$, Bradley Peterson$^{1,2}$, Rob Sharp$^{7,8}$, Michael Childress$^{10,11}$, Geraint Lewis$^{13}$, Brad Tucker$^{7,12}$, Fang Yuan$^{7,11}$, Tim Abbot$^{14}$, Filipe Abdalla$^{15,16}$, Sahar Allam$^{17}$, Aur{\'e}lien Benoit-L{\'e}vy$^{18,15,19}$, Emmanuel Bertin$^{18,19}$, David Brooks$^{15}$, A. Camero~Rosell$^{20,21}$, Matias Carrasco~Kind$^{22,23}$, Jorge Carretero$^{24,25}$, Luiz N.~da Costa$^{20,21}$, Shantanu Desai$^{26,27}$, Thomas Diehl$^{17}$, Tim Eifler$^{28,29}$, David Finley$^{17}$, Brenna Flaugher$^{17}$, Karl Glazebrook$^{30}$, Daniel Gruen$^{31,32}$, Robert Gruendl$^{22,23}$, Gaston Gutierrez$^{17}$, Samuel Hinton$^{6}$, Klaus Honscheid$^{2,33}$, David James$^{14}$, Kyler Kuehn$^{3}$, Nikolav Kuropatkin$^{17}$, Edward Macaulay$^{6}$, M.~A.~G.~Maia$^{20,21}$, Ramon Miquel$^{34,25}$, Ricardo Ogando$^{20,21}$, Andres Plazas$^{29}$, Kevin Riel$^{32}$, Eusebio Sanchez$^{35}$, Basillio Santiago$^{36,20}$, Michael Schubnell$^{37}$, Ignacio Sevilla-Noarbe$^{35}$, R.~C.~Smith$^{14}$, Marcelle Soares-Santos$^{17}$, Flavia Sobreira$^{38,20}$, Eric Suchyta$^{28}$, Molly Swanson$^{23}$, Gregory Tarle$^{37}$, Daniel Thomas$^{39}$, Syed Uddin$^{30}$, Alistair Walker$^{14}$, Bonnie Zhang$^{40}$, The DES Collaboration}
\\
$^{1}$Department of Astronomy, The Ohio State University, Columbus, Ohio 43210, USA \\
$^{2}$Center for Cosmology and Astro-Particle Physics, The Ohio State University, Columbus, OH 43210 \\
$^{3}$Australian Astronomical Observatory, 105 Delhi Rd, North Ryde NSW 2113, Australia \\
$^{4}$Kavli Institute for Cosmology, University of Cambridge, Madingley Road, Cambridge CB3 0HA, UK \\
$^{5}$Institute of Astronomy, University of Cambridge, Madingley Road, Cambridge, CB3 0HA, UK \\
$^{6}$School of Mathematics and Physics, University of Queensland, QLD 4072, Australia \\
$^{7}$CAASTRO: ARC Centre of Excellence for All-sky Astrophysics \\
$^{8}$Research School of Astronomy and Astrophysics, Australian National University, Cotter Rd., Weston ACT 2611, Australia \\
$^{9}$Department of Physics and Astronomy, Curtin University, Kent Street, Bentley, Perth, Western Australia, 6102 \\
$^{10}$School of Physics and Astronomy, University of Southampton, Southampton, SO17 1BJ, UK \\
$^{11}$Research School of Astronomy and Astrophysics, Australian National University, Canberra, ACT 2611, Australia \\
$^{12}$Research School of Astronomy and Astrophysics, Mt. Stromlo Observatory, the Australian National University, Cotter Rd., Canberra ACT 2611, Australia \\
$^{13}$Sydney Institute for Astronomy, School of Physics, A28, The University of Sydney, NSW 2006, Australia \\
$^{14}$Cerro Tololo Inter-American Observatory, National Optical Astronomy Observatory, Casilla 603, La Serena, Chile \\
$^{15}$Department of Physics \& Astronomy, University College London, Gower Street, London, WC1E 6BT, UK \\
$^{16}$Department of Physics and Electronics, Rhodes University, PO Box 94, Grahamstown, 6140, South Africa \\
$^{17}$Fermi National Accelerator Laboratory, P. O. Box 500, Batavia, IL 60510, USA \\
$^{18}$CNRS, UMR 7095, Institut d'Astrophysique de Paris, F-75014, Paris, France \\
$^{19}$Sorbonne Universit{\'e}s, UPMC Univ Paris 06, UMR 7095, Institut d'Astrophysique de Paris, F-75014, Paris, France \\
$^{20}$Laborat{\'o}rio Interinstitucional de e-Astronomia - LIneA, Rua Gal. Jos{\'e} Cristino 77, Rio de Janeiro, RJ - 20921-400, Brazil \\
$^{21}$Observat{\'o}rio Nacional, Rua Gal. Jos{\'e} Cristino 77, Rio de Janeiro, RJ - 20921-400, Brazil \\
$^{22}$Department of Astronomy, University of Illinois, 1002 W. Green Street, Urbana, IL 61801, USA \\
$^{23}$National Center for Supercomputing Applications, 1205 West Clark St., Urbana, IL 61801, USA \\
$^{24}$Institut de Ci{\`e}ncies de l'Espai, IEEC-CSIC, Campus UAB, Carrer de Can Magrans, s/n,  08193 Bellaterra, Barcelona, Spain \\
$^{25}$Institut de F\'{\i}sica d'Altes Energies (IFAE), The Barcelona Institute of Science and Technology, Campus UAB, 08193 Bellaterra (Barcelona) Spain \\
$^{26}$Excellence Cluster Universe, Boltzmannstr.\ 2, 85748 Garching, Germany \\
$^{27}$Faculty of Physics, Ludwig-Maximilians-Universit{\"a}t, Scheinerstr. 1, 81679 Munich, Germany \\
$^{28}$Department of Physics and Astronomy, University of Pennsylvania, Philadelphia, PA 19104, USA \\
$^{29}$Jet Propulsion Laboratory, California Institute of Technology, 4800 Oak Grove Dr., Pasadena, CA 91109, USA \\
$^{30}$Centre for Astrophysics \& Supercomputing, Swinburne University of Technology, Victoria 3122, Australia \\
$^{31}$Kavli Institute for Particle Astrophysics \& Cosmology, P. O. Box 2450, Stanford University, Stanford, CA 94305, USA \\
$^{32}$SLAC National Accelerator Laboratory, Menlo Park, CA 94025, USA \\
$^{33}$Department of Physics, The Ohio State University, Columbus, OH 43210, USA \\
$^{34}$Instituci{\'o} Catalana de Recerca i Estudis Avan{\c{c}}ats, E-08010 Barcelona, Spain \\
$^{35}$Centro de Investigaciones Energ{\'e}ticas, Medioambientales y Tecnol{\'o}gicas (CIEMAT), Madrid, Spain \\
$^{36}$Instituto de F\'{\i}sica, UFRGS, Caixa Postal 15051, Porto Alegre, RS - 91501-970, Brazil \\
$^{37}$Department of Physics, University of Michigan, Ann Arbor, MI 48109, USA \\
$^{38}$ICTP South American Institute for Fundamental Research Instituto de F\'{\i}sica Te{\'o}rica, Universidade Estadual Paulista, S\~ao Paulo, Brazil \\
$^{39}$Institute of Cosmology \& Gravitation, University of Portsmouth, Portsmouth, PO1 3FX, UK \\
$^{40}$The Research School of Astronomy and Astrophysics, Australian National University, ACT 2601, Australia \\
E-mail: mudd@astronomy.ohio-state.edu}

\begin{document}

\maketitle

\begin{abstract}
We present the discovery of a z=0.65 low-ionization broad absorption line (LoBAL) quasar in a post-starburst galaxy in data from the Dark Energy Survey (DES) and spectroscopy from the Australian Dark Energy Survey (OzDES).  LoBAL quasars are a minority of all BALs, and rarer still is that this object also exhibits broad FeII (an FeLoBAL) and Balmer absorption.  This is the first BAL quasar that has signatures of recently truncated star formation, which we estimate ended about 40 Myr ago.  The characteristic signatures of an FeLoBAL require high column densities, which could be explained by the emergence of a young quasar from an early, dust-enshrouded phase, or by clouds compressed by a blast wave.  The age of the starburst component is comparable to estimates of the lifetime of quasars, so if we assume the quasar activity is related to the truncation of the star formation, this object is better explained by the blast wave scenario.
\end{abstract}

\section{Introduction}
\label{sec: intro}
Correlations of the mass of the central supermassive black hole (SMBH) with host galaxy properties such as velocity dispersion \citep{Gebhardt00a, Ferrarese00} suggest that a SMBH's growth is linked to the evolution of the host galaxy through some feedback process (e.g. \citealp{Heckman14}).  The most pronounced phase of SMBH growth is the quasar phase, where most of the spectral energy distribution can be explained by a thin accretion disk of material around the SMBH \citep{Shakura73, Koratkar99}, a hot corona, and a broad line region on larger scales (e.g. \citealp{Peterson97}).  One method of triggering quasar activity is a merger that involves at least one gas-rich galaxy \citep{Sanders88, Silk98, Komossa03, DiMatteo05, Piconcelli10}.  During gas-rich mergers, gas funnels towards the central regions of the galaxies and some fraction accretes onto the central SMBH.  This substantial influx of gas and dust may often obscure the early phases of quasar activity.

Gas-rich mergers also produce a large increase in star formation, up to 100-1000 times the galaxy's quiescent rate.  These rates can quickly exhaust the gas supply, and eventually the star formation rate must return to a lower value.  It is unclear if this is primarily caused by the expulsion of star-forming gas due to the quasar, feedback from the star formation process, or consumption of the gas by star formation and the SMBH.  If the increased star formation rate decreases quickly compared to the total star formation history, the galaxy goes through a post-starburst phase, which is characterized by the strong absorption lines prevalent in A type stars combined with a K type spectrum from an older population \citep{Dressler83, Zabludoff96}.  The absence of stellar features due to shorter-lived O and B type stars would indicate star formation ceased tens to hundreds of Myr ago.  Some post-starburst galaxies also exhibit blueshifted absorption from winds \citep{Tremonti07, Coil11}, and at least some wind-driven outflows from starbursts appear to be delayed by 10 Myr or more after the star formation burst \citep{Sharp10, Ho16}.

A significant fraction of all post-starburst galaxies host quasars \citep{Brotherton99, Brotherton02, Cales13} or some form of lower luminosity active galactic nuclei (AGN).  \citet{Goto06} found that 0.2\% of all galaxies are in a post-starburst state, compared to 4.2\% of quasars having these post-starburst features.  Quasars hosted by post-starburst galaxies typically had intense star formation that ended $10^{8-9}$ years ago.  \citet{Cales13} found that older post-starburst quasars in elliptical galaxies tend to have signs of a recent merger, which suggests that a major merger event fueled both the previous star formation and current quasar activity.  \citet{Tremonti07} and others argue that the presence of blueshifted absorption of a few hundred to a few thousand \kms\textrm{} in some post-starburst quasars is evidence that these objects had a large, galaxy-scale wind $\sim10^{8}$ years ago, although the energy in these winds may not be enough to have quenched star formation \citep{Coil11}.  Similar winds are seen in ongoing starbursts, but these tend to be a factor of a few weaker than in post-starbursts of comparable luminosity \citep{Tremonti07}.  

When the winds from a quasar are especially prominent, they are classified as broad absorption line (BAL) quasars.  BALs are characterized by prominent, blueshifted absorption lines of 2000 \kms\textrm{} or more \citep{Weymann91}.  BALs are present in 20-40\% of all quasars, depending on the selection method \citep{Trump06, Dai08, Urrutia09}.  The majority of BALs only exhibit absorption in high ionization states, such as CIV, and are referred to as HiBALs.  BALs with absorption in lower ionization lines, such as MgII, are referred to as LoBALs.  A small subset of LoBALs also have FeII and/or FeIII absorption and are known as FeLoBALs \citep{Hazard87}.  Rarest of all are the handful of objects with absorption in the Balmer lines \citep{Hall07, Zhang15}.  Using SDSS data, \citet{Trump06} find that HiBALs, LoBALs, and FeLoBALs constitute 26\%, 1.3\%, and 0.3\%, respectively, of their sample of over 16,000 quasars.  In contrast, \citet{Dai08} and \citet{Urrutia09} find BALs are much more common.  When selected with both SDSS and 2MASS to alleviate the bias from reddening, they report 37\%, 32\%, and 32\% of quasars are HiBALs, LoBALs, and FeLoBALs, respectively.  This selection method identifies all LoBALs as FeLoBALs.

FeLoBALs are the most heavily reddened BAL subtype, and the iron features necessitate high column densities (e.g. \citealp{Korista08}).  They can have broad iron emission and absorption from FeIII in addition to FeII, and, in very rare cases, only FeIII \citep{Hall02}.  The absorption troughs are also observed to vary between objects from several distinct, narrow troughs, to blanketing most of the emission from the quasar shortward of the MgII doublet.  Both LoBALs and FeLoBALs also tend to be X-ray faint, further implying that there is a large column density that prevents a direct view of the central source (e.g. \citealp{Mathur95, Green01}).

There remains much debate about the exact nature of FeLoBALs.  With their considerable reddening and high inferred column densities, some argue that they are transitional quasars, moving from a dust-enshrouded star formation phase to an unobscured quasar phase \citep{Voit93, Egami96, Farrah07, Farrah10}.  The highly absorbed FeLoBALs are also more likely to be radio sources, and may be transition objects between radio loud and radio quiet quasars \citep{Becker97}.  Alternatively, \citet{FG12} propose that the absorption is from high density clouds along the line of sight that have been disrupted by a blast wave from the SMBH, rather than a wind pushing out a dusty cocoon.  This would create the absorbers in-situ, allowing them to be either close to the central AGN or farther out in the galaxy but along our line of sight.  The young, dust-enshrouded scenario is less flexible, as the absorbers should be within the central few parsecs.  

We have discovered an FeLoBAL quasar with Balmer absorption and a post-starburst spectrum that was selected using data obtained by the the Dark Energy Survey (DES; \citealp{Flaugher05, Flaugher15}) and the OzDES\footnote{Australian Dark Energy Survey; alternatively, Optical redshifts for DES} collaboration \citep{Yuan15}.  The quasar was found in one of the 10 ``supernova fields'' (3 deg$^{2}$ each, \citealp{Kessler15}) that are monitored to discover Type Ia supernovae.  OzDES obtains approximately monthly spectra of the 10 supernova fields with the AAOmega spectrograph \citep{Smith04, Sharp06} on the 4m Anglo-Australian Telescope (AAT).  Two of its main science goals are measuring redshifts for thousands of host galaxies of Type Ia supernovae discovered with DES photometry and repeatedly observing hundreds of quasars as part of a large-scale reverberation mapping project \citep{King15}.  OzDES also obtains spectra of various other classes of objects, including luminous red galaxies, BAL quasars, and white dwarfs.  

Several of the targets for the DES/OzDES reverberation mapping project are BAL quasars that were selected to monitor their long-term absorption and emission line variability.  Upon stacking several spectra, we discovered that one of these objects, \balnamelong\textrm{} (hereafter \balnameshort), resides in a post-starburst galaxy.  This appears to be the first known BAL quasar in a post-starburst galaxy.  We also note that this is a FeLoBAL with Balmer absorption, making it rare even among BALs, and that it was first chosen as a target candidate from a combination of optical and infrared color cuts described in \citet{Banerji15}, Equation 6.  In Section \ref{sec: obs}, we describe the DES and OzDES observations and accumulate other values from the literature on this unique object.  In Section \ref{sec: analysis}, we characterize both the outflow and model the properties of the host galaxy stellar population using the stacked OzDES spectra.  We  summarize and present our conclusion in Section \ref{sec: conc}.  

\section{Observations}
\label{sec: obs}
All of the spectra of \balnameshort\textrm{} were obtained with the AAT 4m at Siding Spring Observatory as part of the OzDES project.  The double beam fiber-fed spectrograph uses the 580V grating and 385R gratings leading to dispersions of 1\AA/pixel and 1.6\AA/pixel in the blue and red arms, respectively, with the dichroic split at 5700\AA.  The resolution of the spectrograph is $R\sim1400$, and the wavelength range spans 3700-8800\AA.  

\begin{figure}
  \centerline{
    \includegraphics[width=10.0cm]{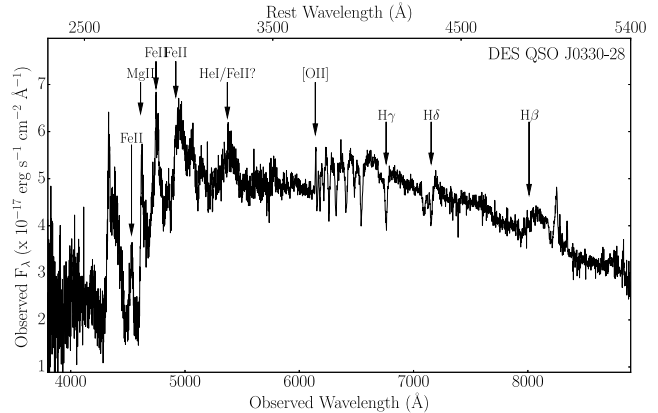}
}
  \caption{Stacked spectrum of \balnameshort\textrm{} at z = 0.65.  The LoBAL features are prominent at wavelengths shorter than the MgII line at rest-frame 2798\AA.  The absorption features around rest-frame 3900\AA\textrm{} are from host galaxy stars.}
  \label{fig: stack_spectrum}
\end{figure}

We present the stacked spectrum in Figure \ref{fig: stack_spectrum} in both the observed and rest frame.  This is a combination of four spectra taken over the course of two years and the combined exposure time is 160 minutes.  We derived the host galaxy redshift of $z$ = 0.65 based on the higher order Balmer lines around rest-wavelength 4000\AA.  There is also a prominent Balmer break shortward of the absorption.  These are the signs of a post-starburst galaxy with recently quenched star formation.  At shorter wavelengths, there is a sharp drop in flux at the rest wavelength of the MgII 2798\AA\textrm{} emission line.  This corresponds to blueshifted absorption out to 5000 \kms\textrm{} from the systemic redshift.  Other absorption troughs in the rest-frame UV correspond to metastable states of Fe II, particularly at 2750\AA, 2880\AA, and 2985\AA.  There may be MgI 2853\AA\textrm{}, but this falls in an FeII absorption trough.  The most common FeIII features are blueward of our spectral coverage.  

\input{table1.tex}

\begin{figure}
  \centerline{
    \includegraphics[width=8.0cm]{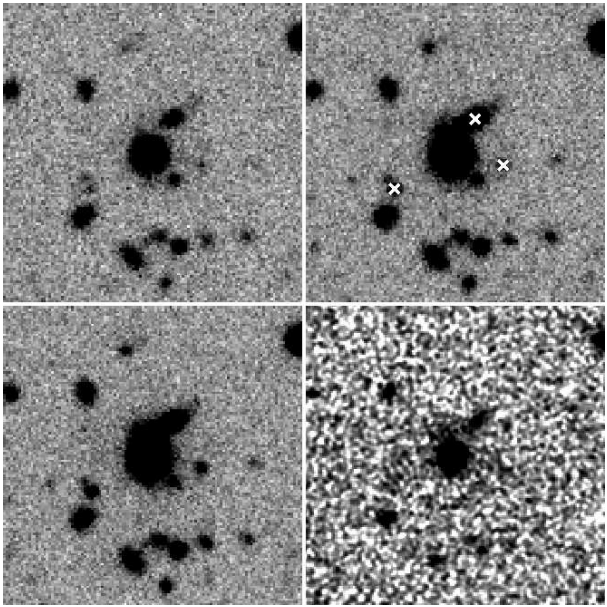}
}
  \caption{DES images of \balnameshort\textrm{} in $g\textrm{ (top left)}, r\textrm{ (top right)}, i\textrm{ (bottom left)},\textrm{ and }K\textrm{ (bottom right)}$.  Each box is $30^{\prime\prime}$ on a side centered on the quasar.  The $g$, $r$, and $i$ images are from DES, and the $K$ band image is from the VISTA VIDEO \citep{Jarvis13} survey.  The three crosses in the $r$\textrm{} image correspond to three sources that have photometric redshifts consistent with \balnameshort\textrm{}, which suggests a merger.}
  \label{fig: DES_images}
\end{figure}

We provide photometry for this object in Table \ref{tab: photo}.  This incorporates \desfilters\textrm{} from DES, \vhsfilters\textrm{} from the VHS survey \citep{McMahon13}, and \emph{W1}, \emph{W2}, \emph{W3}, \emph{W4} from \emph{WISE} \citep{Wright10}.  The DES and WISE magnitudes are calculated using PSF fits, whereas the VHS data use a $2^{\prime\prime}$ aperture.  All magnitudes have been transformed to the AB system.  Both the very red colors (e.g. $r - K$ = 0.86 AB) and spectral shape indicate very substantial reddening, which is quite common with FeLoBALs \citep{Sprayberry92, Hall97}.  The DES $g, r, i$\textrm{} and VISTA $K$\textrm{} images are shown in Figure \ref{fig: DES_images}.  These images show several small objects in the immediate vicinity of the quasar that suggest an interacting or merging system, and three of the objects have photometric redshifts consistent with \balnameshort.  This quasar was also detected as a radio source in the ATLAS survey \citep{Franzen15, Mao12} at 1.474 GHz.  If we extrapolate the ATLAS measurement to 5 GHz with a $\alpha=0.7$, the ratio of rest frame 5 GHz flux density to that at 4400$\AA$ is about two.  This quasar is consequently radio quiet/intermediate under the definition that a ratio less than one is quiet and greater than ten is radio loud.  The result is consistent with the idea that LoBALs may be quasars moving between a radio loud and radio quiet phase and some work suggests that the LoBAL fraction in quasars decreases as a function of radio luminosity \citep{Dai12}.  This object is not detected in archival \emph{Chandra} data, consistent with previous studies that have found FeLoBALs are X-ray faint \citep{Mathur95, Green01}.

\section{Analysis}
\label{sec: analysis}
\subsection{Post-Starburst}
\label{subsec: psb}
\begin{figure}
  \centerline{
    \includegraphics[width=10.0cm]{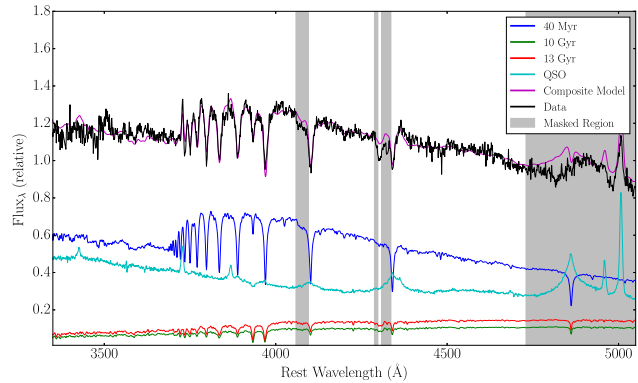}
}
  \caption{The best fit single metallicity model with Z = 0.02$Z_{\odot}$.  The data (black) are fit by a model (magenta) that combines a quasar template (cyan) and three major stellar components: 44\% of the light comes from a younger, recently quenched population with an age of 40 Myr (dark blue), and 24\% comes from older populations of 10 and 13 Gyr (green and red, respectively).  The remainder of the light comes from the quasar.  The masked regions are left out of the fit due to possible broad quasar emission from H$\beta$ and broad absorption in the wings of higher order Balmer lines from the quasar.}
  \label{fig: psb_comp}
\end{figure}

We fit the stacked spectrum with STARLIGHT \citep{CF04, CF05a, CF05b} over the wavelength span not dominated by the FeLoBAL's broad absorption and emission lines (see Figure \ref{fig: psb_comp}).  This corresponds to approximately 3300-4800\AA\textrm{} in the rest frame.  We do not fit to longer wavelengths in order to avoid H$\beta$ contamination.  To account for the quasar component, we created a quasar template from stacked spectra of 10 quasars from the reverberation mapping sample that are most similar in redshift and luminosity to \balnameshort.  We initially ran STARLIGHT over a grid of models supplied by \citet{Bruzual03} that span ages of 1 Myr to 13 Gyr and metallicites from 0.005-2.5$Z_{\odot}$.  The best fit model has approximately 45\% of the light from two young stellar populations of 40 and 55 Myr, 40\% from our quasar template, and the remainder from an older population of 6-7 Gyr.  The metallicity for the varied components is consistent with subsolar to solar.  This fit has \redchi\textrm{} = 0.85.  We also performed fits at single metallicities and found in most instances that between 30-50\% of the light is from 40 and 55 Myr populations and 20-50\% is from the quasar.  These fits had \redchi\textrm{} ranging from 0.9-1.2 and show the relative insensitivity of the population ages to the metallicity.  The strength of the higher order Balmer lines depths do not match perfectly with any age/metallicity combination.  This is likely because of the impact of Balmer absorption in the BAL, and perhaps also some mismatch with the quasar template and this quasar.  For each grid of models, we also fit for the best global extinction and best extinction for each component.  We found the best fit $A_{V}$ was 0-0.04 in most cases.  However, there is a class of single-metallicity models around solar where the best fit has a younger population (5 Myr) with some extinction ($A_{V}$ = 0.37) as the dominant stellar component.   

\subsection{BAL QSO}
\label{subsec: balqso}
There are a number of absorption troughs present at shorter wavelengths than the stellar absorption features in addition to broad absorption associated with some of the Balmer lines.  BAL features are typically described by their balnicity index.  This metric originated in \citet{Weymann91} for HiBALs and the CIV line.  By their definition, a quasar was considered a BAL if it had a balnicity index $BI > 0$.  Later, \citet{Hall02} proposed the absorptive index (AI) as an alternative identifier, which is more sensitive to troughs at lower velocities and likewise identifies BALs with $AI > 0$.  Both BI and AI are integrals over velocity on the blue side of an emission line.  The BI requires the trough to extend at least 3000\kms\textrm{} and drop by at least 10\% of the normalized conitnuum flux.  The AI, however, begins the integral at 0\kms\textrm{} and is more sensitive to lower velocity and weaker troughs.  

\begin{figure}
  \centerline{
    \includegraphics[width=10.0cm]{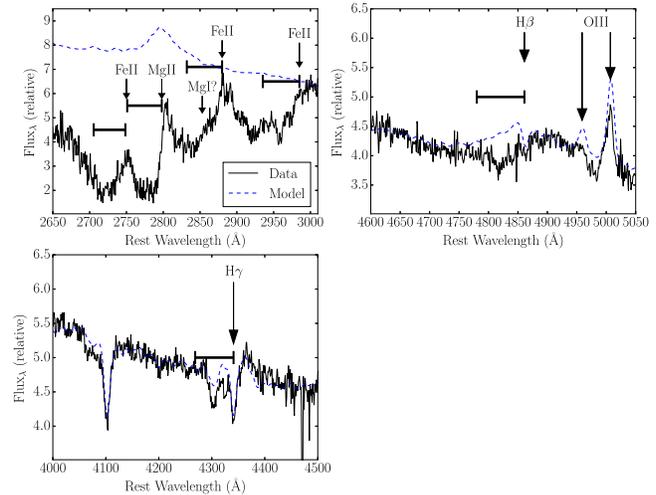}
}
  \caption{Highlight of the BAL troughs.  In each inset, the arrows correspond to the systemic redshift and the horizontal bars correspond to a blueshift velocity of 5000 \kms.  This fits well for MgII and FeII, and there is also Balmer absorption that is consistent with this outflow velocity.  The dotted line is the best fit model for the quasar and stellar components.  The model fits well around the Balmer lines, but vastly overestimates the flux at shorter wavelengths.}
  \label{fig: bal_comp}
\end{figure}

It is difficult to measure these values in FeLoBALs like \balnameshort\textrm{} because these objects have such heavy reddening and the widespread iron absorption/emission makes the continuum poorly defined.  The STARLIGHT fit, partially because it could only fit a narrow wavelength range due to the BAL features, has a best fit $A_{V}$ of 0.04.  However, \balnameshort\textrm{} is clearly highly reddened at shorter wavelengths (see Figure \ref{fig: stack_spectrum}).  To correct for this, we applied various values of $A_{V}$ to our quasar template for an SMC extinction curve \citep{Gordon03} until we found the best fit to the red half of the MgII emission line at 2798\AA.  While no single value gives a satisfactory fit to either the extinction or the continuum, the spectral slope is broadly consistent with $A_{V} = 1-1.5$\textrm{} mag.  This is small given how X-ray faint \citep{Green01} and red many LoBALs are, but is also poorly constrained by the available data.  A likely cause for the difficulty is that there may be partial obscuration; that is, varying amounts of extinction to different regions of the galaxy and quasar emission region.  Without a good continuum fit, we cannot reliably measure AI or BI for this object.  Nevertheless, the velocity spread of the absorption troughs is reasonably clear.  Figure \ref{fig: bal_comp} shows that the MgII absorption spans approximately 5000 \kms\textrm{} before a small rise that is likely due to FeII emission at 2750\AA, which then has its own blueshifted absorption.  The depth and width of the trough means that this object would likely meet the conditions for both AI and BI > 0 for MgII.  

We next compare the velocity extent of the MgII component to other absorption troughs, namely FeII at 2750\AA, 2880\AA, and 2985\AA, and H$\beta$.  Figure \ref{fig: bal_comp} shows in both cases the data are consistent with a similar range of blueshifted absorption, and this suggests a common origin.  

BALs have a large diversity of spectral morphologies, and \balnameshort\textrm{} is most similar to SDSS J112526.13+002901.3 \citep{Hall02} with regard to the approximate shape and strength of the emission and absorption features.  That SDSS quasar has zero balnicity index, but an absorptive index of almost 500\kms.  While we cannot cleanly measure the continuum of \balnameshort, it should also have a nonzero absorptive index. \citet{Hall02} also classified SDSS J112526.13+002901.3 as a many narrow troughs FeLoBAL with HeI absorption.  We do not see evidence for either of these characteristics in \balnameshort.  SDSS J112526.13+002901.3 also does not have the same post-starburst features that make \balnameshort\textrm{} unique.  

We perform a similar analysis to \citet{Hall07} for our H$\beta$ absorption to determine a lower limit column density $N_{H\beta} = 5.2\times10^{14}\textrm{ cm}^{-2}$.  This value is about a factor of 100 smaller than the column density measurement for the \citet{Hall07} FeLoBAL, but likely underestimated for \balnameshort\textrm{} due to the strong unabsorbed galaxy emission at these wavelengths.

\section{Discussion and Conclusion}
\label{sec: conc}
The current picture for quasar evolution in the merger scenario begins with the collision of a dust- and gas-rich galaxy with another system.  Dynamical processes drive material towards the galaxies' centers and fuels star formation and accretion onto the SMBH.  The quasar is initially obscured by the dust, but eventually the material disperses and the quasar becomes easily visible.  FeLoBALs have attracted particular interest because the very high column density absorption is indicative of a substantial outflow, perhaps associated with this transition from obscured to unobscured.  A second scenario proposed by \citet{FG12} is that a blast wave is launched from the quasar that impacts a high density cloud along the line of sight.  This would also create the observed column densities, reddening, and absorption troughs seen in FeLoBALs.  One distinction between these scenarios is in where the absorbing material lies.  For the transition objects, the absorbing material would be around the quasar and in the process of being blown away, whereas for the blast wave model it is possible to impact a cloud on much larger scales than the central few parsecs.

Variability is one way to test the location of the absorbers. The constraints from several variable FeLoBALs (e.g., \citealp{Hall11, McGraw15, Vivek12}) place the absorbing material on the order of a few to tens of parsecs from the central source.  This assumes a cloud-crossing model where the changes arise from an absorber moving across the line of sight.  In contrast, other studies suggest the absorption is on kpc scales \citep{Moe09, Bautista10, Dunn10}.  \citet{Moe09} derived a distance to one outflow of $\sim$3 kpc and, for an assumed covering fraction of 0.2, find that the energy in the BAL outflow is comparable to 1\% of the total luminosity of the quasar.

We found that the obscuration of \balnameshort\textrm{} cannot be fit by a single-screen model.  It is clear that the quasar light is highly reddened at the shortest wavelengths, which implies substantial extinction of the AGN accretion disk.  In contrast, we see little to no extinction of the host galaxy.  The ``transition object'' scenario has this geometry as the young quasar begins to clear out the dust immediately surrounding it to become optically visible.  However, such differential extinction could also arise from a \citet{FG12} blast wave colliding with a dense cloud along our line of sight, either close to the AGN or much farther out in the galaxy.  For the latter case, the quasar can be highly absorbed and the stellar component less so if there is a low covering fraction of the galaxy-wide absorbers.

If we assume the quasar and starburst triggered simultaneously, we can use the starburst age to evaluate which model FeLoBAL scenario is more probable.  Star formation was truncated or quenched in \balnameshort\textrm{} around 50 Myr ago in most of our models.  This time is comparable to estimates for quasar lifetimes of around $10^{7-8}$ years (e.g., \citealp{Yu02, Martini04}) and implies that this FeLoBAL did not turn on recently, which is in conflict with the young quasar scenario.  The FeLoBAL features cannot be due to the same feedback processes that abruptly ended the star formation $\sim$50 Myr ago, as they would have dispersed due to Kelvin-Helmholtz or Rayleigh-Taylor instabilities.  These features are consistent with the \citet{FG12} model in which the absorption is produced by clouds of material that have been compressed by a radiative blast wave.  The key aspect of the blast wave model for \balnameshort\textrm{} is that the blast wave is not tied to a particular evolutionary phase of the quasar.  

We plan to obtain future, higher signal-to-noise ratio spectra over a broader wavelength range to derive better stellar population and reddening parameters.  We will also obtain new spectral epochs as the OzDES program progresses, and we will use these data to search for BAL variability to attempt to measure the distance of the absorber from the central source.

\section{Acknowledgements}
DM would like to gratefully acknowledge the helpful comments by Smita Mathur on a draft of this manuscript.

Funding for the DES Projects has been provided by the U.S. Department of Energy, the U.S. National Science Foundation, the Ministry of Science and Education of Spain, 
the Science and Technology Facilities Council of the United Kingdom, the Higher Education Funding Council for England, the National Center for Supercomputing 
Applications at the University of Illinois at Urbana-Champaign, the Kavli Institute of Cosmological Physics at the University of Chicago, 
the Center for Cosmology and Astro-Particle Physics at the Ohio State University,
the Mitchell Institute for Fundamental Physics and Astronomy at Texas A\&M University, Financiadora de Estudos e Projetos, 
Funda{\c c}{\~a}o Carlos Chagas Filho de Amparo {\`a} Pesquisa do Estado do Rio de Janeiro, Conselho Nacional de Desenvolvimento Cient{\'i}fico e Tecnol{\'o}gico and 
the Minist{\'e}rio da Ci{\^e}ncia, Tecnologia e Inova{\c c}{\~a}o, the Deutsche Forschungsgemeinschaft and the Collaborating Institutions in the Dark Energy Survey. 

The Collaborating Institutions are Argonne National Laboratory, the University of California at Santa Cruz, the University of Cambridge, Centro de Investigaciones Energ{\'e}ticas, 
Medioambientales y Tecnol{\'o}gicas-Madrid, the University of Chicago, University College London, the DES-Brazil Consortium, the University of Edinburgh, 
the Eidgen{\"o}ssische Technische Hochschule (ETH) Z{\"u}rich, 
Fermi National Accelerator Laboratory, the University of Illinois at Urbana-Champaign, the Institut de Ci{\`e}ncies de l'Espai (IEEC/CSIC), 
the Institut de F{\'i}sica d'Altes Energies, Lawrence Berkeley National Laboratory, the Ludwig-Maximilians Universit{\"a}t M{\"u}nchen and the associated Excellence Cluster Universe, 
the University of Michigan, the National Optical Astronomy Observatory, the University of Nottingham, The Ohio State University, the University of Pennsylvania, the University of Portsmouth, 
SLAC National Accelerator Laboratory, Stanford University, the University of Sussex, Texas A\&M University, and the OzDES Membership Consortium.

The DES data management system is supported by the National Science Foundation under Grant Number AST-1138766.
The DES participants from Spanish institutions are partially supported by MINECO under grants AYA2012-39559, ESP2013-48274, FPA2013-47986, and Centro de Excelencia Severo Ochoa SEV-2012-0234.
Research leading to these results has received funding from the European Research Council under the European Union’s Seventh Framework Programme (FP7/2007-2013) including ERC grant agreements 
 240672, 291329, and 306478.

The data in this paper were based on observations obtained at the Australian Astronomical Observatory.  

Part of this research was conducted by the Australian Research Council Centre of Excellence for All-sky Astrophysics (CAASTRO), through project number CE110001020.

\bibliographystyle{mn2e}
\bibliography{bal_psb_references}

\end{document}

%% file: table1.tex
\begin{table}
\begin{minipage}{8in}
\caption{\balnameshort\textrm{} Photometry}
\begin{tabular}{lrrcrrrrrr}
\hline
\hline
{Band Name} & Cent. Wave & {Magnitude (Error)} \\
\hline
\emph{g} & 5720\AA & 20.11 (0.02) \\
\emph{r} & 6590\AA & 19.31 (0.02)\\
\emph{i} & 7890\AA & 19.02 (0.02)\\
\emph{z} & 9760\AA & 18.94 (0.02)\\
\emph{Y} & 1${\mu}$m & 19.00 (0.02)\\
\emph{J} & 1.25${\mu}$m & 18.89 (0.04)\\
\emph{H} & 1.65${\mu}$m & 18.77 (0.05)\\
\emph{K} & 2.15${\mu}$m & 18.45 (0.05)\\
\emph{W1} & 3.4${\mu}$m & 17.64 (0.03)\\
\emph{W2} & 4.6${\mu}$m & 17.01 (0.03)\\
\emph{W3} & 12${\mu}$m & 15.92 (0.06)\\
\emph{W4} & 22${\mu}$m & 15.01 (0.22)\\
ATLAS & 1.474 GHz & $256^{a} (20)$ \\
\hline
\hline
\label{tab: photo}
\end{tabular}
\end{minipage}
Photometry for \balnameshort\textrm{} taken from DES for \desfilters, VHS for \vhsfilters\textrm{} \citep{McMahon13}, and WISE for W1-W4 \citep{Wright10}.  The DES data are PSF magnitudes obtained from the coadd of the first year of observations.  All magnitudes are given in the AB system aside from the radio data from ATLAS \citep{Franzen15, Mao12}.\\
$^{a}$This is in $\mu$janksy rather than magnitudes.\\
\end{table}